\documentclass[1p,number,times,sort&compress]{elsarticle}
\usepackage[dvipsnames,table,xcdraw]{xcolor}
\usepackage[normalem]{ulem} 
\usepackage{cancel}
\usepackage{graphicx} 
\usepackage{subfigure} 
\usepackage{chngcntr} 
\usepackage{amssymb} 
\usepackage{amsmath} 
\usepackage{mathrsfs}
\usepackage{epstopdf}
\usepackage{float} 
\usepackage{booktabs} 
\usepackage{microtype}
\usepackage[font=footnotesize]{caption}
\usepackage[colorlinks,breaklinks]{hyperref}
 
\makeatletter
\renewcommand{\p@subfigure}{\thefigure}
\makeatother

\begin{document} 
 

\title{Building Three-Dimensional Differentiable Manifolds Numerically II:
       Limitations}

\author[cass,ccom]{Lee Lindblom\corref{cor1}}
\ead{llindblom@ucsd.edu}
\author[HTWBerlin]{Oliver Rinne}
\ead{oliver.rinne@htw-berlin.de}
\cortext[cor1]{Corresponding author}

\address[cass]{Center
  for Astrophysics and Space Sciences, University of California at San
  Diego,\\ 9500 Gilman Drive, La Jolla, CA 92093, USA}
\address[ccom]{Center for Computational Mathematics,
  University of California at San Diego,
  \\ 9500 Gilman Drive, La Jolla, CA 92093, USA}
\address[HTWBerlin]{Faculty 4, HTW Berlin -- University of Applied Sciences,
Treskowallee 8, 10318 Berlin, Germany} 

\date{\today}
 
\maketitle

\section{Summary}
\label{s:Summary}

Methods were developed in Ref.~\cite{Lindblom2022} for constructing
reference metrics (and from them differentiable structures) on
three-dimensional manifolds with topologies specified by suitable
triangulations.  This note generalizes those methods by expanding the
class of suitable triangulations, significantly increasing the number
of manifolds to which these methods apply.  These new results show
that this expanded class of triangulations is still a small subset of
all possible triangulations.  This demonstrates that fundamental
changes to these methods are needed to further expand the collection
of manifolds on which differentiable structures can be constructed
numerically.

\section{Fixing the Dihedral Angles}
\label{s:FixingDihedralAngles}

The method for constructing reference metrics in
Ref.~\cite{Lindblom2022} begins with the construction of a flat metric
in the neighborhood of each vertex of a multicube structure, which can
be obtained from a triangulation of that manifold.  These flat metrics
are then combined using partition of unity functions to produce a
global $C^0$ metric, and then smoothed to $C^1$ by a sequence of
additional steps described in Ref.~\cite{Lindblom2022}.  These flat
metrics are constructed by fixing the dihedral angles of each cube
edge.  The simple method used in Ref.~\cite{Lindblom2022} fixes those
dihedral angles to be $2\pi/K$, where $K$ is the number of cube edges
that intersect along a particular edge.  This choice ensures the sum
of the dihedral angles around each edge is $2\pi$, the condition
needed to avoid a conical singularity there.  This uniform dihedral
angle condition severely limits the class of multicube structures on
which it can be applied.  This simple condition is replaced here with
more complicated but less restrictive conditions.

The basic adjustable parameters that determine these flat metrics are
the dihedral angles, $\psi_{A\{\alpha\beta\}}$, where the index
$A\in\langle1,...,N_\mathrm{cubes}\rangle$ labels the cubes in the
multicube structure and $\{\alpha\beta\}\in\bigl\langle\{-x-y\},
\{-x+y\}, \{-x-z\}, \{-x+z\}, \{+x-y\}, \{+x+y\}, \{+x-z\}, \{+x+z\},
\{-y-z\}, \{-y+z\}, \{+y-z\}, \{+y+z\}\bigr\rangle$ labels the edge
formed by the intersection of the $\{\alpha\}$ and
$\{\beta\}\in\bigl\langle\{-x\}, \{+x\}, \{-y\}, \{+y\}, \{-z\},
\{+z\}\bigr\rangle$ faces of that cube.  Each cube has 12 edges so
there are a total of $12N_\mathrm{cubes}$ dihedral angle parameters
needed to determine the vertex centered flat metrics.

The sum of the dihedral angles, $\psi_{A\{\alpha\beta\}}$, from the
cubes that intersect along an edge must equal $2\pi$ to avoid a conical
singularity along that edge.  This constraint can be written
explicitly:
\begin{equation}
  0=\mathscr{C}_{A\{\alpha\beta\}}
  \equiv 2\pi - \!\!\! \sum_{A'\{\alpha'\beta'\}}\psi_{A'\{\alpha'\beta'\}}\,,
  \label{e:EdgeConstraints}
\end{equation}
where the sum is over all the edges that intersect along edge
$A\{\alpha\beta\}$.  Many of these $12N_\mathrm{cubes}$ constraints
are redundant, but for simplicity all are enforced in the numerical
analysis here.

\begin{figure}[!hbt] 
  \begin{center}  
    \includegraphics[width=0.35\textwidth]{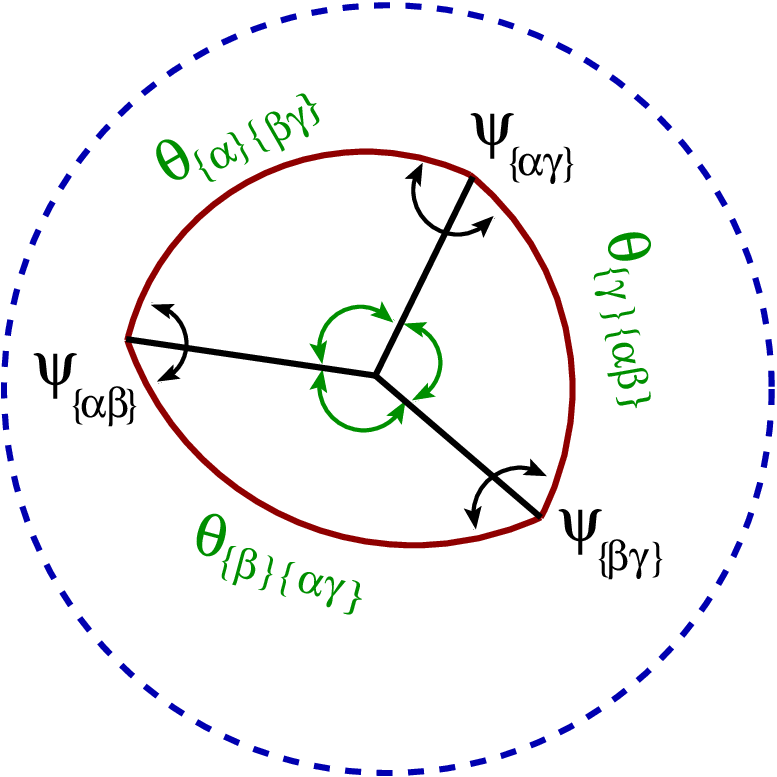}
    \caption{Figure shows the intersection between the corner of a
      cubic region and a small sphere centered on one of the vertices
      of that cube.  This sphere is depicted as the dashed (blue)
      curve; the intersection of this cubic region with the sphere is
      a spherical triangle shown as solid (red) curves; the solid
      (black) straight lines are the edges of the cube.  The dihedral
      angles $\psi_{\{\alpha\beta\}}$ between the cube faces are also
      the angles of this spherical triangle.  The vertex angles,
      $\theta_{\{\alpha\}\{\beta\gamma\}}$, are the angles between the
      edges of the cube, and are also the arc lengths of the sides of
      this spherical triangle.
      \label{f:ThreeDWedge} }
  \end{center} 
\end{figure}

Another set of important angles are the vertex angles between the
edges of the cube, see Fig.~\ref{f:ThreeDWedge}. The notation
$\theta_{A\{\gamma\}\{\alpha\beta\}}$ is used for these angles, where
the $A$ index labels the cube, $\{\gamma\}$ one of the cube faces, and
$\{\alpha\beta\}$ the edge that intersects $\{\gamma\}$ at the
$\{\alpha\beta\gamma\}$ vertex.  These
$\theta_{A\{\gamma\}\{\alpha\beta\}}$ are the angles between vectors
tangent to the $\{\alpha\gamma\}$ and the $\{\beta\gamma\}$ edges.
There are three vertex angles $\theta_{A\{\gamma\}\{\alpha\beta\}}$
associated with each vertex, so 24 for each cube and
$24N_\mathrm{cubes}$ total for the multicube structure.  The law of
cosines from spherical trigonometry gives the relationship between a
vertex angle $\theta_{A\{\gamma\}\{\alpha\beta\}}$ and the dihedral
angles $\psi_{A\{\alpha\beta\}}$ associated with the cube edges that
intersect at that vertex:
\begin{eqnarray}
\cos \theta_{\{\gamma\}\{\alpha\beta\}} = \frac{\cos \psi_{\{\alpha\beta\}} -
  \cos \psi_{\{\alpha\gamma\}} \cos \psi_{\{\beta\gamma\}}}
     {\sin \psi_{\{\alpha\gamma\}} \sin \psi_{\{\beta\gamma\}}}\,.
     \label{e:LawOfCosines}
\end{eqnarray}
The $\theta_{A\{\gamma\}\{\alpha\beta\}}$ can therefore be considered
functions of the $\psi_{A\{\alpha\beta\}}$.
\vspace{-0.75cm}

Consider two cube faces, $A\{\alpha\}$ and $A'\{\alpha'\}$ that are
identified in the multicube structure.  The intrinsic metrics
associated with these cube faces can only be continuous across the
interface between cubes if the vertex angles
$\theta_{A\{\alpha\}\{\beta\gamma\}}$ are the same as the
corresponding angles $\theta_{A'\{\alpha'\}\{\beta'\gamma'\}}$ on the
identified face.  These additional constraints on the dihedral angles
$\psi_{A\{\alpha\beta\}}$ can be written:
\begin{eqnarray}
0=\mathscr{C}_{A\{\alpha\}\{\beta\gamma\}} \equiv \cos
\theta_{A\{\alpha\}\{\beta\gamma\}} - \cos
\theta_{A'\{\alpha'\}\{\beta'\gamma'\}}\,.
\label{e:VertexAngleConstraints}
\end{eqnarray}
Any interface with $\mathscr{C}_{A\{\alpha\}\{\beta\gamma\}}\neq 0$
has a metric discontinuity and consequently a curvature singularity at
that interface. Half of these vertex angle constrains are redundant,
but for simplicity all are enforced in the numerical analysis here.

In any multicube structure the $N_\mathrm{edges}$ independent
constraints in Eq.~(\ref{e:EdgeConstraints}) (where $N_\mathrm{edges}$
are the number of independent edges in the multicube structure) and
the $12N_\mathrm{cubes}$ independent constraints in
Eq.~(\ref{e:VertexAngleConstraints}) must be satisfied by the
$12N_\mathrm{cubes}$ dihedral angle parameters.  Since there are more
constraints than freely specifiable parameters, we expect that many
(most) multicube structures will not admit solutions to all the
constraints. When solutions do exist we expect they are likely to be
unique in most cases.  In Ref.~\cite{Lindblom2022} a relatively small
collection of multicube structures were found that admit uniform
dihedral angle solutions to these constraints.  Solutions that do not
satisfy the uniform dihedral angle condition are found here for a
wider class of manifolds.

The vertex angle constraints, Eq.~(\ref{e:VertexAngleConstraints}),
are very nonlinear, and general analytic solutions are not known.
More general solutions can be found numerically, however, by finding
the minima of the combined constraint norm, $||\mathscr{C}||$, defined
by
\begin{equation}
  ||\mathscr{C}||^2 = \sum_{A\{\alpha\beta\}} \mathscr{C}_{A\{\alpha\beta\}}^2
  +\sum_{A\{\gamma\}\{\alpha\beta\}}\mathscr{C}_{A\{\gamma\}\{\alpha\beta\}}^2\,,
\label{e:ConstraintNorm}
\end{equation}
where the sums are over the $12N_\mathrm{cubes}$ edge and the
$24N_\mathrm{cubes}$ vertex angle constraints defined in
Eqs.~(\ref{e:EdgeConstraints}) and (\ref{e:VertexAngleConstraints}).
This norm is a function of the dihedral angles, $||\mathscr{C}||^2 =
||\mathscr{C} (\psi_{A\{\alpha\beta\}})||^2$ that is bounded below by
zero, so a minimum always exists.  If $\min ||\mathscr{C}||^2=0$ then
all the constraints are satisfied. If $\min ||\mathscr{C}||^2\neq 0$
then the constraints are not satisfied and it is not possible to build
a non-singular $C^0$ metric on that multicube structure in this way.

The numerical search for a minimum of $||\mathscr{C}||^2$ was started
by setting initial guesses for $\psi_{A\{\alpha\beta\}}$ to their
uniform dihedral angle values: $\psi_{A\{\alpha\beta\}}=
2\pi/K_{A\{\alpha\beta\}}$, where $K_{A\{\alpha\beta\}}$ is the number
of cube edges that intersect edge $A\{\alpha\beta\}$.  The numerical
search for a minimum was carried out using the
Broyden--Fletcher--Goldfarb--Shanno (BFGS) algorithm
\cite[p. 136]{Nocedal2006} in the Python library
\texttt{scipy.optimize}.  Any numerical minimum with
$||\mathscr{C}||\leq 10^{-12}$ was considered to be a good numerical
solution to all the constraints, while any minimum with
$||\mathscr{C}||> 10^{-12}$ was rejected. Our interest is finding
solutions to these constraints that can be used to construct reference
metrics on these manifolds.  Our numerical searches for solutions were
concentrated near the uniform dihedral angle state, because only
relatively undistorted multicube structures are useful to us as
computational domains for solving partial differential equations
numerically.

Constraint-satisfying dihedral angles $\psi_{A\{\alpha\beta\}}$ were
searched for numerically on the 744 multicube structures constructed
from the triangulations having eight or fewer tetrahedra included in
the Regina~\cite{Regina} catalog of compact orientable
three-dimensional manifolds.  Table~\ref{t:uniform_list} lists the 23
manifolds from this search that satisfy all the constraints as well as
the uniform dihedral angle condition.  Table~\ref{t:optimized_list}
lists 80 additional manifolds that admit non-uniform dihedral angle
solutions with $||\mathscr{C}||\leq 10^{-12}$.  The manifold names
used in these tables are those from the Regina~\cite{Regina} catalog.
These tables also list the number of multicube regions,
$N_\mathrm{cubes}$, and the maximum number of edges,
$K_\mathrm{max}=\max K_{A\{\alpha\beta\}}$, that overlap in each
multicube structure.  Table~\ref{t:optimized_list} also includes two
parameters, $\min \mathcal{A}_{A\{\alpha\beta\gamma\}}$ and $\min \det
g^{-1}_{A\{\alpha\beta\gamma\}}$, that measure how distorted the
constraint-satisfying dihedral angles make each multicube region.  The
quantity $\mathcal{A}_{A\{\alpha\beta\gamma\}}$ is the solid angle
subtended by the cube at the $A\{\alpha\beta\gamma\}$ vertex (i.e. the
area of the spherical triangle in Fig.~\ref{f:ThreeDWedge}).  This
solid angle would equal $\tfrac{\pi}{2}$ in an un-distorted cube, so
$\tfrac{2}{\pi}\min \mathcal{A}_{A\{\alpha\beta\gamma\}}$ is a good
measure of the maximum distortion in a multicube structure.  The
quantity $\det g^{-1}_{A\{\alpha\beta\gamma\}}$ represents the
determinant of the inverse $C^0$ metric constructed in
Ref.~\cite{Lindblom2022} from the dihedral angles, evaluated at the
vertex ${A\{\alpha\beta\gamma\}}$.  This determinant would equal one
in an un-distorted cube, so $\min \det
g^{-1}_{A\{\alpha\beta\gamma\}}$ is another good measure of the
maximum distortion in a multicube structure.  Multicube structures
with $\tfrac{2}{\pi}\min \mathcal{A}_{A\{\alpha\beta\gamma\}} <
10^{-6}$ or $\min \det g^{-1}_{A\{\alpha\beta\gamma\}} < 10^{-6}$ were
excluded from the list in Table~\ref{t:optimized_list}.

These results show that only a small fraction, $(23+80)/744\approx
0.138$, of the multicube structures constructed from eight or fewer
triangulations in the Regina~\cite{Regina} catalog allow dihedral
angles that satisfy all the constraints. These include only three
manifolds constructed from eight tetrahedra, and an even smaller
fraction is expected for the manifolds based on triangulations with
more tetrahedra.  These results reveal that the methods developed in
Ref.~\cite{Lindblom2022}, including the generalizations presented
here, are unfortunately quite limited in their ability to construct
reference metrics on all the manifolds based on the triangulations in
the Regina~\cite{Regina} catalog.
\begin{table}[!h] 
    \scriptsize \renewcommand{\arraystretch}{1.4}
\begin{center}
  \caption{Multicube Structures Admitting Uniform Dihedral Angles.
    The manifold names used here are those from the Regina catalog.
    Those names are explained in detail in the documentation to the
    Regina cagalog~\cite{Regina} and also in
    Ref.~\cite{Lindblom2022}.  \label{t:uniform_list}}
  \begin{tabular}{l c c| l c c}  
    \midrule
    Manifold & $N_\mathrm{cubes}$ & $K_\mathrm{max}$
    & Manifold & $N_\mathrm{cubes}$ & $K_\mathrm{max}$
    \\
    \midrule
    L(5,2) & 4 & 4
      & SFS[RP2/n2:(2,1)(2,-1)] & 24 & 6 \\
    L(8,3) & 8 & 4
      & SFS[S2:(2,1)(2,1)(2,-1)] & 8 & 4 \\ 
    L(10,3) & 12 & 6
      & SFS[S2:(2,1)(2,1)(3,-2)] & 12 & 6 \\
    L(12,5) & 12 & 6
      & SFS[S2:(2,1)(2,1)(4,-3)] & 16 & 8 \\
    L(16,7) & 16 & 8
      & SFS[S2:(2,1)(2,1)(5,-4)] & 20 & 10 \\ 
    L(20,9) & 20 & 10
      & SFS[S2:(2,1)(2,1)(6,-5)] & 24 & 12 \\
    L(24,11) & 24 & 12
      & SFS[S2:(2,1)(2,1)(7,-6)] & 28  & 14 \\
    L(28,13) & 28 & 14
      & SFS[S2:(2,1)(2,1)(8,-7)] & 32 & 16 \\
    L(32,15) & 32 & 16
      & SFS[S2:(2,1)(3,1)(5,-4)] & 20 & 5 \\
    T$\times$S1 & 24 & 6
      & SFS[S2:(2,1)(3,2)(3,-1)] & 20 & 5 \\
    KB/n2$\times\!\!\sim$S1 &24 & 6
      & SFS[S2:(2,1)(4,1)(4,-3)] & 24 & 6 \\
    & &
      & SFS[S2:(3,1)(3,1)(3,-2)] & 24 & 6 \\
    \bottomrule
  \end{tabular}
\end{center}
\end{table}
\begin{table}[!t]  
    \scriptsize
  \renewcommand{\arraystretch}{1.4}
\begin{center}
  \caption{Multicube Structures Admitting Non-uniform Dihedral Angles
    With $||\mathscr{C}||\leq 10^{-12}$. The quantities
    $\tfrac{2}{\pi}\min \mathcal A$ and $\min \mathrm{det}\,g^{-1}$
    (defined in the text) measure distortion of the multicube
    structure, with $\tfrac{2}{\pi}\min \mathcal A=\min
    \mathrm{det}\,g^{-1} = 1$ in an undistorted structure.  Bold face
    entries also admit non-singular $C^1$ metrics.
    \label{t:optimized_list}}
    \begin{tabular}{l c c c c |l c c c c}  
    \midrule
    Manifold & $N_\mathrm{cubes}$ & $K_\mathrm{max}$ 
    &  $\frac{2}{\pi}\,{\min \mathcal A}$ & $\min \mathrm{det}\,g^{-1}$
    & Manifold & $N_\mathrm{cubes}$ & $K_\mathrm{max}$
    &  $\frac{2}{\pi}\,{\min \mathcal A}$ & $\min \mathrm{det}\,g^{-1}$\\
    \midrule
    \bf L(7,2)  & \bf 8  & \bf 5  & \bf 0.652  & \bf 0.380 
      & L(50,19) & 24 & 7 & 0.117 & 0.026 \\
    \bf L(11,3) & \bf 12 & \bf 6 & \bf 0.424   & \bf 0.202
      & L(55,21) & 24 & 6 & 0.121 & 0.027 \\
    L(13,3) & 16 & 8 & 0.169 & 0.036
      & \bf SFS[S2:(2,1)(2,1)(2,1)] & \bf 16 & \bf 5 & \bf 0.332 & \bf 0.495 \\
    \bf{L(13,5)} & \bf{12} & \bf{5} & \bf{0.598} & \bf{0.500}
      & SFS[S2:(2,1)(2,1)(2,3)] & 20 & 7 & 0.228 & 0.125 \\
    L(14,3) & 16 & 8 & 0.122 & 0.016 
      & SFS[S2:(2,1)(2,1)(2,5)] & 24 & 9 & 0.198 & 0.125 \\
    L(15,4) & 16 & 6 & 0.360 & 0.112
      & SFS[S2:(2,1)(2,1)(2,7)] & 28 & 11 & 0.122 & 0.073 \\
    L(17,4) & 20 & 8 & 0.188 & 0.021
      & \bf SFS[S2:(2,1)(2,1)(3,-1)] & \bf 16 & \bf 5 & \bf 0.332 & \bf 0.495 \\
    L(17,5) & 16 & 7 & 0.358 & 0.198
      & \bf SFS[S2:(2,1)(2,1)(3,1)] & \bf 20 & \bf 7 & \bf 0.218 & \bf 0.216 \\
    L(18,5) & 16 & 6 & 0.348 & 0.123
      & \bf SFS[S2:(2,1)(2,1)(3,2)] & \bf 20 & \bf 6 & \bf 0.173  & \bf 0.232 \\
    L(19,4) & 20 & 8 & 0.101 & 0.007
      & SFS[S2:(2,1)(2,1)(3,4) & 24 & 8 & 0.122 & 0.031 \\
    \bf L(19,7) & \bf 16 & \bf 7 & \bf 0.384 & \bf 0.361
      & SFS[S2:(2,1)(2,1)(3,5) & 24 & 8 & 0.088 & 0.050 \\
    \bf L(21,8) & \bf 16 & \bf 6 & \bf 0.344 & \bf 0.379
      & SFS[S2:(2,1)(2,1)(3,7) & 28 & 10 & 0.075 & 0.027 \\
    L(22,5) & 20 & 8 & 0.132 & 0.010
      & \bf SFS[S2:(2,1)(2,1)(4,-1)] & \bf 20 & \bf 6 & \bf 0.332 & \bf 0.247 \\
    L(23,5) & 20 & 8 & 0.104 & 0.006
      &  SFS[S2:(2,1)(2,1)(4,1)] & 24 & 9 & 0.095 & 0.012 \\
    \bf L(23,7) & \bf 20 & \bf 9 & \bf 0.246 & \bf 0.128
      &  SFS[S2:(2,1)(2,1)(4,3)] & 24 & 7 & 0.070 & 0.057 \\
    \bf L(24,7) & \bf 20 & \bf 8 & \bf 0.270 & \bf 0.093
      & \bf SFS[S2:(2,1)(2,1)(5,-3)] & \bf 20 & \bf 7 & \bf 0.143 & \bf 0.182 \\
    L(25,7) & 20 & 6 & 0.252 & 0.055
      & SFS[S2:(2,1)(2,1)(5,-2)] & 20 & 6 & 0.242 & 0.197 \\
    \bf L(25,9) & \bf 20 & \bf 9 & \bf 0.268 & \bf 0.204
      & SFS[S2:(2,1)(2,1)(5,2)] & 24 & 8 & 0.055 & 0.043 \\
    L(26,7) & 20 & 7 & 0.236 & 0.034
      & SFS[S2:(2,1)(2,1)(5,3)] & 24 & 7 & 0.045 & 0.071 \\
    L(27,8) & 20 & 8 & 0.234 & 0.075
      & SFS[S2:(2,1)(2,1)(5,4)] & 28 & 9 & 0.025 & 0.005  \\
    L(29,8) & 20 & 6 & 0.216 & 0.032
      & SFS[S2:(2,1)(2,1)(7,-5)] & 24 & 9 & 0.079 & 0.092 \\
    L(29,9) & 24 & 11 & 0.191 & 0.091 
     & SFS[S2:(2,1)(2,1)(7,-4)] & 24 & 8 & 0.138 & 0.095 \\
    L(29,12) & 20 & 7 & 0.236 & 0.099
     & SFS[S2:(2,1)(2,1)(7,-3)] & 24 & 7 & 0.135 & 0.012 \\
    L(30,7) & 24 & 9 & 0.121 & 0.005
     & SFS[S2:(2,1)(2,1)(7,-2)] & 24 & 8 & 0.151 & 0.046 \\
    L(30,11) & 20 & 8 & 0.232 & 0.179
     & SFS[S2:(2,1)(2,1)(8,-5)] & 24 & 8 & 0.112 & 0.095 \\
    L(31,11) & 24 & 11 & 0.206 & 0.130
     & SFS[S2:(2,1)(2,1)(8,-3)] & 24 & 7 & 0.132 & 0.035 \\
    L(31,12) & 20 & 7 & 0.214 & 0.140
     & SFS[S2:(2,1)(2,1)(9,-7)] & 28 & 11 & 0.049 & 0.056 \\
    L(33,7) & 24 & 8 & 0.073 & 0.002
     & SFS[S2:(2,1)(2,1)(9,-5)] & 28 & 9 & 0.065 & 0.0006 \\
    L(33,10) & 24 & 10 & 0.196 & 0.063 
     & SFS[S2:(2,1)(2,1)(10,-7)] & 28 & 10 & 0.067 & 0.042 \\
    L(34,9) & 24 & 9 & 0.179 & 0.128
     & SFS[S2:(2,1)(2,1)(12,-5)] & 28 & 8 & 0.005 & 0.00002 \\
    L(34,13) & 20 & 6 & 0.200 & 0.121
      & \bf SFS[S2:(2,1)(3,1)(3,-2)] & \bf 16 & \bf 5 & \bf 0.334 & \bf 0.500 \\
    L(35,11) & 28 & 13 & 0.156 & 0.067
      & SFS[S2:(2,1)(3,1)(3,-1)] & 20 & 6 & 0.170 & 0.138 \\
    L(36,11) & 24 & 10 & 0.182 & 0.058
      & SFS[S2:(2,1)(3,1)(3,1)] & 24 & 7 & 0.147 & 0.069 \\
    L(37,10) & 24 & 8 & 0.155 & 0.020
      & SFS[S2:(2,1)(3,1)(3,2)] & 24 & 7 & 0.104 & 0.055 \\
    L(39,14) & 24 & 10 & 0.176 & 0.102
      & \bf SFS[S2:(2,1)(3,1)(4,-3)] & \bf 20 & \bf 6 & \bf 0.167 & \bf 0.250 \\
    L(41,12) & 24 & 8 & 0.149 & 0.014
      & SFS[S2:(2,1)(3,1)(4,-1)] & 24 & 7 & 0.193 & 0.74 \\
    L(42,13) & 28 & 12 & 0.155 & 0.044
      & \bf SFS[S2:(2,1)(3,1)(5,-3)] & \bf 24 & \bf 7 & \bf 0.164 & \bf 0.228\\
    L(43,15) & 32 & 15 & 0.138 & 0.066
      & SFS[S2:(2,1)(3,2)(4,-3)] & 24 & 6 & 0.091 & 0.022 \\
    L(44,13) & 24 & 8 & 0.122 & 0.009
      & SFS[S2:(2,1)(3,2)(4,-1)] & 24 & 7 & 0.151 & 0.029 \\
    L(48,17) & 28 & 12 & 0.139 & 0.061 
      & SFS[S2:(2,1)(3,2)(5,-2)] & 24 & 6 & 0.067 & 0.082 \\
    \bottomrule
  \end{tabular}
  \end{center}
\end{table}

\vspace{-0.75cm}

\section{Discussion}
\label{s:Discussion}

Given a set of constraint-satisfying dihedral angles
$\psi_{A\{\alpha\beta\}}$, a global $C^0$ reference metric can be
constructed in a straightforward way using the methods developed in
Ref.~\cite{Lindblom2022}.  These $C^0$ reference metrics determine a
basic $C^1$ differentiable structure on those manifolds.  Smoother
differentiable structures are needed, however, to allow global
solutions to second-order equations like Einstein's gravitational
field equation.  Methods for transforming the $C^0$ reference metrics
to $C^1$ (or smoother via Ricci flow) are also given in
Ref.~\cite{Lindblom2022}.  Those methods were used here successfully
to construct $C^1$ reference metrics for the 23 manifolds listed in
Table~\ref{t:uniform_list} and the 17 manifolds displayed in boldface
in Table~\ref{t:optimized_list}.  The constraint-satisfying dihedral
angles make the remaining 63 non-boldface manifolds in
Table~\ref{t:optimized_list} so distorted that non-singular $C^1$
metrics could not be constructed in this way.  Considerable effort was
expended in various attempts to find improved methods that could
produce non-singular $C^1$ metrics in those cases, but all those
efforts failed.  Only $23+17$ of the 744 triangulations from the
Regina~\cite{Regina} catalog produced $C^1$ reference metrics that
could be used to solve Einstein's equation on those manifolds.  These
results demonstrate that the methods developed in
Ref.~\cite{Lindblom2022}, and generalized here, are very limited.
Expanding the class of manifolds on which useful $C^1$ reference
metrics can be constructed will probably require new methods for
transforming the triangulations on those manifolds into ones that
cover those manifolds more ``uniformly'' and so admit less distorted
multicube structures.

\section*{Acknowledgments}

Research was supported by NSF grant 2012857 to the
University of California at San Diego.

\vspace{0.2cm} \bibliographystyle{model1-num-names}
\bibliography{../References/References}

\begin{thebibliography}{3}
\expandafter\ifx\csname natexlab\endcsname\relax\def\natexlab#1{#1}\fi
\providecommand{\bibinfo}[2]{#2}
\ifx\xfnm\relax \def\xfnm[#1]{\unskip,\space#1}\fi
\bibitem[{Lindblom et~al.(2022)Lindblom, Rinne, and Taylor}]{Lindblom2022}
\bibinfo{author}{L.~Lindblom}, \bibinfo{author}{O.~Rinne},
  \bibinfo{author}{N.~W. Taylor},
\newblock \bibinfo{title}{Building differentiable three-dimensional manifolds
  numerically},
\newblock \bibinfo{journal}{J.\ Comput.\ Phys.} \bibinfo{volume}{410}
  (\bibinfo{year}{2022}) \bibinfo{pages}{110957}.
\bibitem[{Nocedal and Wright(2006)}]{Nocedal2006}
\bibinfo{author}{J.~Nocedal}, \bibinfo{author}{S.~J. Wright},
  \bibinfo{title}{Numerical Optimization}, \bibinfo{publisher}{Springer},
  \bibinfo{year}{2006}.
\bibitem[{Burton et~al.(2021)Burton, Budney, Pettersson et~al.}]{Regina}
\bibinfo{author}{B.~A. Burton}, \bibinfo{author}{R.~Budney},
  \bibinfo{author}{W.~Pettersson}, et~al., \bibinfo{title}{Regina: Software for
  low-dimensional topology}, \bibinfo{howpublished}{{\tt http://\allowbreak
  regina-normal.\allowbreak github.\allowbreak io/}},
  \bibinfo{year}{1999--2021}.

\end{thebibliography}
\end{document}